# Long Range Anticorrelations and Non-Gaussian Behavior of a Leaky Faucet

T.J.P. Penna*, P.M.C. de Oliveira

*Instituto de Física, Universidade Federal Fluminense,*
*Av. Litorânea, s/n, 24210-340 Niterói, RJ, Brazil*

J.C. Sartorelli, W.M. Gonçalves and R.D. Pinto

*Instituto de Física, Universidade de São Paulo, C.P. 20516, 01452-990 São Paulo, SP, Brazil*

(July 21, 1995)

We find that intervals between successive drops from a leaky faucet display scale-invariant, long-range anticorrelations characterized by the same exponents of heart beat-to-beat intervals of healthy subjects. This behavior is also confirmed by numerical simulations on lattice and it is faucet-width- and flow-rate-independent. The histogram for the drop intervals is also well described by a Lévy distribution with the same index for both histograms of healthy and diseased subjects. This additional result corroborates the evidence for similarities between leaky faucets and healthy hearts underlying dynamics.

Very recently, long-range power-law correlations have been reported in a wide variety of systems as DNA sequences [1], stock market fluctuations [2], literary pieces [3] and heart beat intervals [4,5]. Due to this power-law behavior it is possible to characterize these diverse phenomena by critical exponents and, through the latter, to identify similarities between the systems, which have not been noticed before. In this Letter, we present a novel example of this connection. We measured experimentally the exponents from time series of drops in leaky faucets and we found the same values as obtained by Peng *et al.* [4] for heartbeat time series. Furthermore, we find that the statistics of this process is also described by Lévy distribution. Scale invariance and non-Gaussian behaviors were confirmed by a theoretical model for two-dimensional drops.

Leaky faucets are dynamical systems presenting complex behavior in drop-to-drop interval time series. It has been confirmed by experimental data [6–9] and through numerical simulations [10]. The most relevant parameter in this case is the flow rate. In fig. 1(a,b) we present two experimental time series and the return maps constructed plotting the time interval $B(n)$ between drop $n$ and drop $n + 1$ (we adopt the same notation as ref. [4]). The experimental setup is described in ref. [9]. Each set of experimental data consists of 8192 drops (8 $K$drops, hereafter). We measured fourteen sets at different flow rates.

Recently, a numerical procedure to simulate dripping faucets was introduced [10]. This model is based on the near and next-nearest neighbor Ising model on a square lattice, where a spin up (down) represents the fluid (air). The earth gravitation is imposed as a magnetic field varying uniformly in the vertical direction, and the dynamics is mass-conservative. The return maps obtained from this model are quite similar to those measured on thin faucets [8,9] (inner diameter<1mm). From numerical simulations we have obtained time series with up to 256 $K$drops and 40 values of flow rate and faucet width. Since we can represent the air-fluid system by Boolean variables, the bit-handling techniques for time and memory saving [11] can be used allowing the implementation of our code on microcomputers. Our simulations were carried out in a 486-DX2 microcomputer.

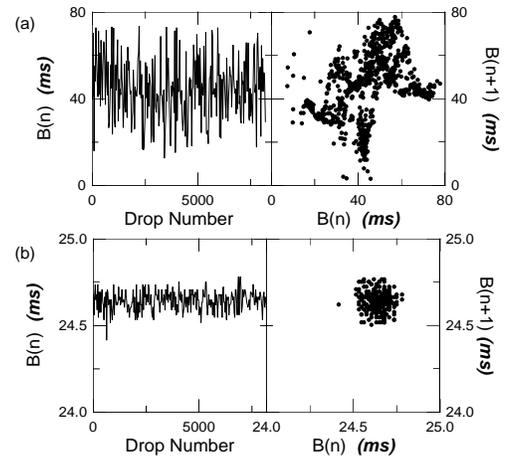

FIG. 1. The drop-to-drop interval $B(n)$, in *ms* units, for different flow rates: (a) 25 drops/s, (b) 40 drops/s (time series and return maps).

In all the cases we have studied, the time series are similar to the ones presented in fig. 1 and to electrocardiograms of healthy subjects [4,12]. Electrocardiograms of patients with dilated cardiomyopathy are smoother than the ones presented here. We can observe the complex pattern of fluctuations emerging in all cases. In the leaky faucet time series, the fluctuations are also related to competition. In the heart, parasympathetic stimulation decreases the firing rate of pacemaker cells whereas sympathetic stimulation acts in the opposite direction. Analogously, in a drop, cohesive forces which create surface tension, which tends to decrease the dripping rate, and gravity are the competing forces. This competition has a nonlinear character and is the natural candidate to explain this complex mechanism. The similarity in the underlying dynamics of these systems is that this tug-of-war is controlled by the surface-volume ratio. This statement is confirmed since this behavior has been found even in the simplified model for two-dimensional leaky faucets, in which no kinetic energy terms (besides the thermal Metropolis energy) were added.

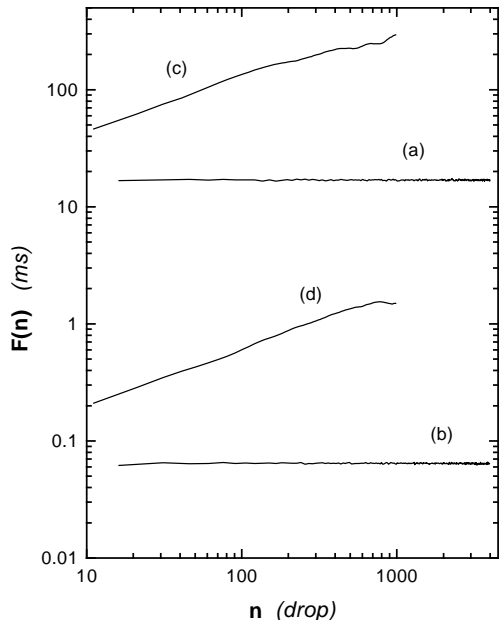

FIG. 2. Log-log plot of $F(n)$ vs $n$ for those time series presented in fig.1. From regression we have found respectively $\alpha = 0.001$ (a); -0.001 (b). Curves (c) and (d) means to a random ordering in time of the events of original time series (a) and (b), respectively.

A quantity frequently used to characterize sequences such as those shown in fig. 1 is the mean fluctuation function $F(n)$, defined as

$$F(n) = \overline{|B(n' + n) - B(n')|} \qquad (1)$$

where the bar denotes average over all values of $n'$ (for a complete description of characterizing sequences as time series and random walks, see refs. [5,13]). If the sequence is a random walk, or the correlations are local, then $F(n) \sim n^{1/2}$ and if no characteristic length exists, $F(n) \sim n^{\alpha}$, with $\alpha \neq 1/2$. Although this power-law behavior is well defined for infinite sequences, it is possible to estimate and minimize the finite size effects [14,15]. The log-log plots of $F(n)$ vs $n$ for the data of fig. 1 are presented in fig. 2. We took 50 different samples of 8 $K$drops, where the sample size is 1024 in order to reduce the fluctuations in the exponent $\alpha$. For the 54 sequences studied we found $\alpha$ ranging between $-0.08$ and $0.09$. Let us stress here that this result is found to be *faucet-width and flow-rate-independent.* If the fluctuactions decay algebraically, the correlation function is also described by a power law $(C(n) \sim (1/n)^{\gamma})$ [5]. The exponents are not independent, since

$$\alpha = \frac{2 - \gamma}{2}. \qquad (2)$$

To our knowledge, this power-law behavior of the correlation function was not reported before, hence it is a novel result concerning the leaky faucet dynamics. In addition, this is the first report of a physical system (besides the healthy heart) presenting anticorrelations. It was widely believed that the correlation function should decay exponentially. According to our results, leaky faucets (and healthy hearts) works at a critical regime, instead of a chaotic one. Our results have greater precision than those of ref. [4], due mainly to difficulty in controlling the patients during long time as required. Hence, leaky faucets seem to be good laboratories to test this kind of studies since they have no physiological constraints. In addition, since we have measured flow rates as low as 7.7 drops/s, i.e., 8 $K$drops in less than twenty minutes, long time instabilities are irrelevant in the range of sequences. We also show, in fig.2c and 2d, two curves corresponding to a deliberated prepared random ordering of interdrop increments, defined as $I(n) = B(n+1) - B(n)$. For these new series, we found $\alpha = 0.5$, which caracterize an uncorrelated sequence.

Another fact reinforcing the analogy between healthy hearts and leaky faucets concerns the histograms of interdrop increments $I(n)$. In fig. 3 we reproduce the best fit of a Lévy stable distribution [16]:

$$P(I, \psi, \delta) = \frac{1}{\pi} \int_0^{\infty} \exp(-\delta q^{\psi}) \cos(qI) dq \qquad (3)$$

for the set of experimental data presented in fig. 1(b). The data were fitted using the recently introduced technique of generalized simulated annealing [17] minimizing the squared error between experimental data and predicted values [2]. The values of $\psi$ were found to range between 1.66 and 1.85, for diverse experimental and numerical data. The Lorentzian distribution is a special case of eq.(3) with $\psi = 1$ and the Gaussian distribution corresponds to $\psi = 2$. For both healthy and diseased hearts one has $\psi = 1.7$ [4]. This agreement strongly suggests that correlations are produced by the same underlying dynamics in human hearts and leaky faucets. Another remarkable similarity is the better fit to positive values of $I$ than to negative ones, exactly as for human hearts [4]. It is important to stress that its non-Gaussian behavior is independent of the aforementioned results concerning long range correlations, because only these latter depend on the ordering of the events (compare fig.2a and 2b with 2c and 2d). *Therefore this feature should be interpreted as an additional evidence for similarities between heartbeats and leaky faucets dynamics.* While in ref. [4], the authors suggest that the slow decay of Lévy distributions may be of physiological importance, we do not understand why this behavior appears on leaky faucets. It is also remarkable that these behaviors had been also detected in the theoretical model. Whereas the attractors of real leaky faucets dynamics are strongly dependent on several parameters as viscosity, diameter and shape of the faucet, etc. [18], the exponents remains the same of the simplified two-dimensional model. This results point out this scale-invariant behavior is inherent to the dynamics.

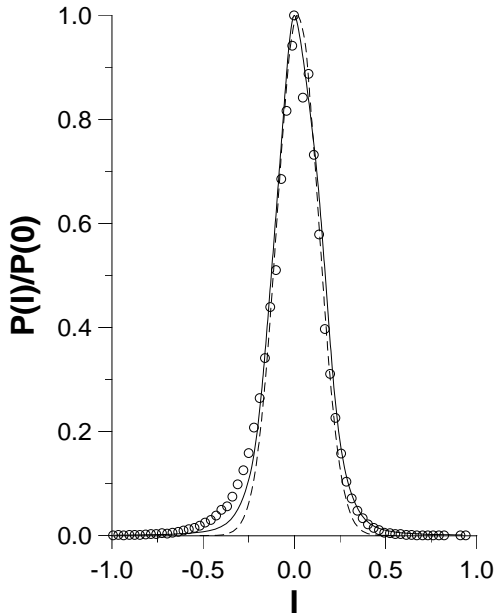

FIG. 3. Histogram of $I(n)$ (open circles) and respective best fits for Gaussian (dotted line) and Lévy (solid line) distributions. The parameters for the best fit Lévy distribution are $\psi = 1.73$ and $\gamma = 0.26$. Frequently the histograms present three peaks: the largest one which mean value vanishes, and two symmetric others ( the one corresponding to large drops followed by small ones and the other corresponding to the inverse event). We select to show here a situation where only one peak appears, hence the fit is more reliable.

We also performed standard spectral analysis on our sequences. Fig.4 is a log-log plot of the power spectra $S(f)$, the square of the Fourier transform amplitudes for $I(n)$. The power law form

$$S(f) \sim (1/f)^\beta \qquad (4)$$

confirms that we have a long-range correlated sequence. The value $\beta = -1$ indicates a non-trivial correlation at all time scales. The relation $\beta = 2\alpha - 1$ is observed with a great accuracy ($\beta = 1.05$), although the dispersion in $\beta$ is remarkably larger than in $\alpha$.

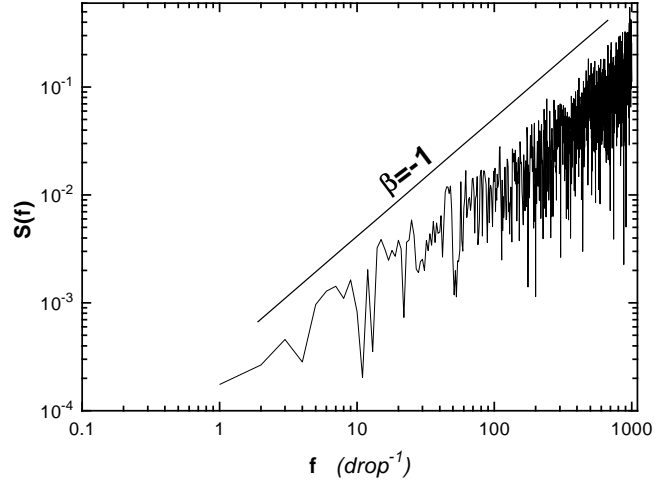

FIG. 4. Power spectra $S(f)$ for the interval increments for the time series presented in fig.1. A straight line corresponding to $\beta = -1$ curve is presented for comparison.

In summary, we have found that time series from leaky faucets display scale-invariant and long range correlations. The exponents of the fluctuaction function and power spectra are numerically the same as the ones from healthy hearts. It is well know that many systems in nature can display this type of scaling behavior. However, since the histograms of drop-to-drop intervals are also well described by Lévy statistics, with the same index as healthy and diseased hearts, we can suspect that the dynamics of these systems is governed by the same mechanism triggered by the surface-volume ratio. Even a simplified model for leaky faucets [10] reflects these properties, therefore this mechanism is robust to increasing of complexity. Therefore, it is not surprising, from an adaptive point of view, that the healthy heart behaves as a leaky faucet and the diseased one does not. Futhermore, to our knowledge, this is the first theoretical model reflecting these properties. Hence, it can be useful, due to its simplicity, in order to get a better understanding of the underlying dynamics. The presence of long-range correlations in biological systems is frequently associated to adaptive behavior. Clearly, leaky faucets do not belong to the class of systems presenting evolution but we believe that the present results *do not discard* the suggestion of healthy heart adaptive behavior. The suppression of an excessive mode-locking due to lack of a characteristic time scale is a considerable advantage of healthy hearts [4,12]. The fact that simple hydrodynamical systems, such as leaky faucets, present long-range anticorrelations, suggests that some similar mechanism may have been important, in the evolutive process, for the heart beating.

## ACKNOWLEDGMENTS

We are grateful to D. Stauffer and S.L.A. de Queiroz for suggestions improving the manuscript. Financial support from the Brazilian Agencies CNPq, CAPES, FAPESP and FINEP is gratefully acknowledged.